\documentclass[12pt,preprint]{aastex}
\begin{document}
\title{Cataclysmic Variables from SDSS. VIII. The Final Year (2007-2008)\footnote{Based on 
observations obtained with the Sloan Digital Sky Survey and with the
 Apache Point
Observatory (APO) 3.5m telescope, which are owned and operated by the
Astrophysical Research Consortium (ARC)}}

\author{Paula Szkody\altaffilmark{2},
Scott F. Anderson\altaffilmark{2},
Keira Brooks\altaffilmark{2},
Boris T. G\"ansicke\altaffilmark{3},
Martin Kronberg\altaffilmark{2},
Thomas Riecken\altaffilmark{2},
Nicholas P. Ross\altaffilmark{4},
Gary D. Schmidt\altaffilmark{5},
Donald P. Schneider\altaffilmark{6},
Marcel A. Ag\"ueros\altaffilmark{7},
Ada N. Gomez-Moran\altaffilmark{8,9},
Gillian R. Knapp\altaffilmark{10},
Matthias R. Schreiber\altaffilmark{11},
Axel D. Schwope\altaffilmark{8}}

\altaffiltext{2}{Department of Astronomy, University of Washington, Box 351580,
Seattle, WA 98195; szkody@astro.washington.edu}
\altaffiltext{3}{Department of Physics, University of Warwick, Coventry CV4 7AL, UK}
\altaffiltext{4}{Lawrence Berkeley National Lab, 1 Cyclotron Road, Berkeley, CA 92420}
\altaffiltext{5}{The University of Arizona, Steward Observatory, Tucson, AZ 85721; currently at NSF}
\altaffiltext{6}{Department of Astronomy and Astrophysics, 525 Davey Laboratory, Pennsylvania State University, University Park, PA 16802}
\altaffiltext{7}{Department of Astronomy, Columbia University, 550 West 120th Street, New York, NY 10027}
\altaffiltext{8}{Leinbiz-Institut fuer Astrophysick Potsdam (AIP), An der Sternwarte 16, 14482 Potsdam, Germany}
\altaffiltext{9}{CNRS, Observatoire Astronomique de Strasbourg, 11 Rue de la Universite, 67000 Strasbourg, France}
\altaffiltext{10}{Princeton University Observatory, Peyton Hall, Princeton NJ 08544}
\altaffiltext{11}{Universidad de Valparaiso, Departamento de Fisica y Astronomia, Chile}

\begin{abstract}
This paper completes the series of cataclysmic variables (CVs) identified from
the Sloan Digital Sky Survey I/II. The coordinates, magnitudes and spectra
of 33 CVs are presented. Among the 33 are eight systems known previous to SDSS
 (CT Ser,
DO Leo, HK Leo, IR Com, V849 Her, V405 Peg, PG1230+226 and HS0943+1404), as
well as nine objects recently found through various photometric surveys.
Among the systems identified since the SDSS are
two polar candidates, two
intermediate polar candidates and one candidate for containing a pulsating
white dwarf. Our followup data have confirmed a polar candidate from Paper VII
and determined tentative periods for three of the newly identified CVs. 
A complete
summary table of the 285 CVs with spectra from SDSS I/II is presented as well 
as a link to an online table of all known CVs from both photometry and 
spectroscopy that will continue to be updated as future data appear.
\end{abstract}


\section{Introduction}
Since the beginning of the Sloan Digital Sky Survey (SDSS; York et al. 2000), we have analyzed
the available spectra to produce a listing of cataclysmic
variables (CVs) in the growing database each calendar year. This paper 
completes the listing for
the last year of the primary SDSS I-II surveys and the data available in the final SDSS I-II
public release (DR7, Adelman-McCarthy et al. 2009, \footnote{Data are
available online from http://www.sdss.org}). The previous listings of
CVs are available in Papers I-VII (Szkody et al. 2002, 2003, 2004, 2005, 2006, 
2007, 2009).
As in the previous papers, the CVs include dwarf novae, novalikes, and
objects with highly magnetic white dwarfs (polars and intermediate polars). 
These types of CVs are
reviewed in detail in Warner (1995). 

The value of the SDSS-selected objects is that they provide a large sample
of CVs with spectra of medium resolution ($\sim$3\AA) and relatively large
wavelength coverage (3800-9200\AA), with accurate coordinates, finding
charts and photometry available for all objects. This has led to
many followup observations by a multitude of groups that resulted in 
the determination of over a hundred orbital periods. With this significant
amount of information,  
the viability of CV population models could be explored. Prior to  DR7,
G\"ansicke et al. (2009) used the available results for 116 systems
to show that the increased depth of SDSS provided a new window on the
period distribution. The periods from the SDSS CVs were much closer
to the predictions of population models (Howell, Nelson \& Rappaport 1995)
than previous surveys with brighter limits that better sampled long
period systems. The final 33 objects in this paper provide a small (10\%)
increment to the total number of SDSS CVs, but complete the work of the last 
decade and present
several interesting systems for future study. In
addition to the new objects, we also include followup polarimetry on
a polar from Paper VII (SDSSJ093839.25+534403.8).  

\section{Observations and Reductions}

The specific details concerning the SDSS survey can be found in Pier et al. 
(2003), Gunn et al.
(1998, 2006); Lupton, Gunn, \& Szalay (1999);
 Hogg et al. (2001);  Lupton et al. (2001); Ivezic et al. (2004); 
 Smith et al. (2002); Tucker et al.
(2006); Padmanabhan et al. (2008). The methods of detecting the CVs are 
stated in Szkody
et al. (2002) and in the publications on the selection algorithms 
(Stoughton et al. 2002, Richards
et al. 2002). As a very brief summary, the objects in the imaging data obtained in
$u,g,r,i,z$ filters (Fukugita et al. 1996) are selected for spectral fibers through the use of
established  color algorithms, with
galaxies and quasars receiving the bulk of the fibers for spectra.
Fortunately, the colors of quasars and hot objects cover many of the CVs
and so most of the acquired spectra come from QSO candidates (see Richards
et al. (2002) for a description of the QSO selection algorithm and G\"ansicke
et al. (2009) for details of the QSO color selection overlap with CVs). Software programs
as well as sporadic eye searches are then used to identify objects with 
zero-redshifts and
Balmer emission/absorption lines that could be CVs. Additionally,
discrepancies in magnitude betweeen the photometry and spectroscopy can
be an indication of an object that has transitioned from outburst to
quiescence or from low to high states of mass transfer. Thus, the selection
is not complete but from comparisons of known CVs from past catalogs (Ritter
\& Kolb 2003) in areas covered by the
SDSS, comparison of our manual searches to computer findings, and
comparison of CV colors to that of QSO's (G\"ansicke et al. 2009), we estimate
that the completeness of blue CVs with i $\le$ 19.1 mag is roughly similar
to that for ultraviolet excess quasars ($\sim$90\%; Schneider et al. 2007).

Table 1 lists the 33 objects in DR7 from 2007 January 1 to the conclusion of
SDSS II in 2008 June. As in the previous papers, this table
shows plate, fiber, modified Julian Date (MJD) of the spectra, along
with the coordinates in equinox J2000.0 (truncated to the last digit)
with an accuracy of 0.10 arcsec. Magnitudes and colors are obtained using
the point spread function (PSF) photometry, with no correction for
reddening. For the rest of this paper, we will simplify the SDSS names
to SDSSJhhmm (hours and minutes of RA) except for SDSS0745, which needs
the first digits of declination as there are two objects with similar
abbreviated RAs.

Followup observations were obtained for a few objects using the
APO 3.5m telescope and the Dual Imaging Spectrograph (DIS) with high
resolution gratings (2\AA) and a 1.5 arcsec slit. One set of
spectra was also accomplished for the possible polar presented
in Paper VII (SDSSJ0938). A summary of these spectral followup observations
is given in Table 2. Spectropolarimetry of SDSSJ0938 was also
obtained on the 2.3m Bok telescope with the CCD Spectropolarimeter (SPOL)
 on 2008 October 29 (4800 sec) and 2009 April 26 (3600 sec), and on the 1.5m 
Kuiper telescope on 2009 Jan 30, 31 (2400 sec
each night) and Feb 1 (1600 sec). 

The calibrations for flux and wavelength for the APO data, as well as
 the line measurements,
were accomplished with standard
IRAF \footnote{{IRAF (Image
 Reduction and Analysis
Facility) is distributed by the National Optical Astronomy Observatories, which
are operated by AURA,
Inc., under cooperative agreement with the National Science Foundation.}}
routines. The 
``e'' (centroid) routine in the
IRAF $\it{splot}$ package was used to measure the SDSS spectra for 
line centers, equivalent widths and fluxes of the Balmer and helium emission lines.
Lines were measured several times to estimate an error on the measured velocities.
The fluxes and equivalent widths are listed in Table 3. 
The solutions for the radial velocity curves for the time-resolved
spectra were obtained by using
 a least squares fit of a sine curve to the velocities
to determine
 $\gamma$ (systemic velocity), 
K (semi-amplitude), P (orbital period), and $T_{0}$ (the UT time of red
to blue crossing of the systemic velocity). A Monte Carlo method was used
to generate the errors on $\gamma$ and K. These solutions are listed in
Table 4 along with the standard deviation ($\sigma$) of the sine fit to the
data points. Since the APO observation times are typically allotted in
half-nights, the periods are not well established. The period uncertainties are
typically 10\% based on differences between the H$\beta$ and H$\alpha$ lines
and should only be used as a basis for classification of an object as
under or over the period gap that exists in the distribution of CVs between 2-3 hrs.

\section{Results}

Figure 1 shows the SDSS spectra of the 33 objects in Table 1. Details
on several classes of systems are presented below.

\subsection{Previously Known Systems}

There are 8 objects in Table 1 that were known as CVs prior to the SDSS.
SDSSJ1545 is Nova Ser 1948 (CT Ser) which has been studied by
Ringwald et al. (2005) who determined an orbital period of 0.195 days.
Four objects were identified as blue sources in the Palomar-Green
survey (Green, Schmidt \& Liebert 1986). SDSSJ1040 is PG1030+155 (DO Leo)
which Abbott et al. (1990) identified as an eclipsing novalike variable with
an orbital period of 5.63 hr. The SDSS spectrum presented in Figure 1 has
a greater strength of \ion{He}{2} than shown in Abbott et al. (1990). 
The ratio of \ion{He}{2}/H$\beta$ in the SDSS spectrum is
1.4 (Table 3) while this ratio is 0.33 in Abbott et al. (1990). SDSSJ1117 is PG1114+187 
(HK Leo), identified as a pre-CV with an orbital period of 1.76 days 
by Hillwig et al. (2000). SDSSJ1232 is PG1230+226, 
which has no further information available. The fourth object, SDSSJ1635, 
is PG1633+115 (V849 Her), and was studied photometrically by
Missalt \& Shafter (1995) who determined a period near 203 minutes but could not sort out aliases with their data. The Hamburg
Quasar survey (Hagen et al. 1995) first found HS0943+1404 which is SDSSJ0946. Rodriguez-Gil
et al. (2005) identified this variable as an intermediate polar with a spin
period of 69 min and an orbital period near 4.2 hr. The last two objects
were found as counterparts to bright objects in the ROSAT survey. IR Com
(SDSSJ1239) was discovered by Richter et al. (1995) and later found to be an 
eclipsing dwarf nova with an orbital period of 125.3 min (Wenzel et al. 1995);
further detailed photometry was accomplished by Feline et al. (2005).
V405 Peg is SDSSJ2309, discovered by Schwope et al. (2002), and
recently studied in detail by Thorstensen et al. (2009), who found it
to have an unusually low accretion rate for a system with an orbital period of 4.26 hr
and suggested that it could be a system in hibernation. Their spectra show states of some accretion (strong emission lines) and states of lower accretion (weak emission). The SDSS spectrum
in Figure 1 appears similar to those of slightly enhanced accretion.  

Recently, Wils et al. (2010) performed a correlation of
the {\it{GALEX}}, Catalina Sky Survey (CSS; Drake et al. 2009) and SDSS photometry as well as a 
search for multiple SDSS
photometric observations that showed variablity, resulting in a set
of CV candidates. Included in this group of likely dwarf novae are
SDSSJ0732,  SDSSJ0756+30, SDSSJ0912, SDSSJ0928, SDSSJ1005, SDSSJ1610,
SDSSJ1625 and SDSSJ1120. Olech et al. (2011) recently determined an orbital
period of 131.2 min for SDSSJ1625 following an outburst of this system.
Two further systems (SDSSJ1055 and SDSSJ1122) 
were identified as CV candidates by the CSS 
and appear on the CSS web page of 
CVs \footnote{http://nesssi.cacr.caltech.edu/catalina.AllCV.html} 
(Drake et al. 2009). SDSSJ1122 was followed up by vsnet during an
outburst (Maehara 2010a,b;
Kato 2010), with superhumps at a period of 65.4 minutes reported.

\subsection{Dwarf Novae}

Dwarf novae are recognized by their outbursts, revealed by several
magnitude increases in photometric measurements as well as by the
spectral changes from Balmer emission to absorption lines during the
outburst. The different times of SDSS photometry and spectroscopy
allow some dwarf novae to be identified. Past United States Naval Observatory
 and Digitized Sky Survey catalogs and
ongoing Catalina Sky Survey (CSS) photometry also provide chances to detect outbursts.
Wils et al. (2010) and Drake et al. (2009) 
used these methods to provide nine dwarf nova candidates. Comments on the
SDSS spectra and our followup observations for several of these systems
are given below.

SDSSJ0928 has strikingly different SDSS photometric and spectroscopic magnitudes 
(the spectrum
in Table 1 was obtained
at outburst) and SDSSJ1120 has two SDSS spectra, one at outburst and
one at quiescence. 

The spectrum of SDSSJ1055
shown in Figure 1 is at outburst, while the SDSS photometry
reports a $g$ mag of 19.2. An APO spectrum obtained at quiescence (Table 2)
is shown in Figure 2. The visibility of a sharp peak component in 
the lines indicates a prominent hot spot may contribute to the light
of this system.

{\it SDSS0912}. The pronounced doubling of the Balmer lines in this object
(Figure 1) indicates a relatively high inclination. 
Time-resolved spectra over 2.3 hrs of this system
during quiescence produced radial velocity curves (Figure 3) 
indicating a period near 2 hrs. Table 4 lists the best fits for
the H$\beta$ and H$\alpha$ lines, which involve a relatively low K
semi-amplitude. However, the relatively poor quality of
the fit requires a longer data string to produce a reliable solution.

\subsection{Polars}

Polars are identified spectroscopically either from the prescence of
strong \ion{He}{2} together with large amplitude radial velocity curves
for these lines or from cyclotron harmonics (Wickramasinghe \& Ferrario 2000). 
The former is usually present for
systems in high states of accretion, while cyclotron humps are most
visible for systems with low accretion (e.g. Schmidt et al. 2005). In both cases, the observation of
circular polarization provides the confirmation of a polar.
Figure 1 shows the spectra of our three prime candidates for polars, SDSSJ1005, SDSSJ1207
and SDSSJ1344. The first two have the strong \ion{He}{2} lines of
a high state, while SDSSJ1344 reveals a single cyclotron hump. It is
possible SDSS1005 may be an intermediate polar (IP), as its spectrum is
similar to the identified IP SDSSJ0946 (Figure 1). Polarization observations will
determine the correct identification.  The followup observations of
SDSSJ1344 and a polar candidate from Paper VII (SDSSJ0938) are discussed below.  

{\it SDSSJ0938}. In Paper VII, SDSSJ0938 was identified as a $g$=19.15 mag polar candidate, with
the spectrum showing strong \ion{He}{2}. We have subsequently obtained
spectropolarimetry when the system was at a similar magnitude (2008 Oct 29)
and also on 4 nights in 2009 Jan-Apr when the system was in a low
state (about a magnitude fainter). Time-resolved APO spectra over 4.2 hrs were 
also obtained in 2009 Feb during the faint state.

The results of the spectropolarimetry are presented in Figure 4. The
circular polarization was measured to be weak but significant and always
showed positive values. There are several puzzling aspects to these
data. The presence of circular polarization confirms that this is
a system with a magnetic white dwarf. As the polarization is present
throughout the optical spectrum, the field must be relatively strong,
and the lack of cyclotron harmonics rules out a low mass accretion system.
However, it is not clear why the polarization is so much smaller than
in normal polars and why the value does not change when the system
goes into the low state and the emission lines disappear. 

The time-resolved APO spectra provided few clues. Whereas 7 spectra
over 70 min obtained during the high state in 2008 (Paper VII)
showed a smooth variation during this interval, the 4.2 hrs during
the low state showed only a constant velocity with a standard deviation
of 19 km s$^{-1}$. Only the
H$\alpha$ line was strong enough to measure.
Further data
when the system re-enters a higher state are needed to resolve the
physical characteristics of this system.

{\it SDSSJ1344}. While the Balmer emission lines of SDSSJ1344 are
relatively weak, the time-resolved APO spectra were able to show
a large radial velocity variation during the 107 min of observation.
Figure 5 shows the radial velocity curves while Table 4 lists the
best fits. The large K amplitude confirms the likely polar nature,
but polarimetry will be required to obtain the field strength
as the single cyclotron hump visible in Figure 1 does not provide
enough information to establish which harmonic is viewed.
 
\subsection{Nova-likes with \ion{He}{2}}

Besides the previously identified IP (SDSSJ0946), the nova CT Ser (SDSSJ1545),  the eclipsing nova-like DO Leo (SDSSJ1040), and the outbursting dwarf nova
(SDSSJ1055), the other objects in Figure 1 that have significant \ion{He}{2}
emission are 
SDSSJ0756+08, SDSSJ1122, 
 SDSSJ1232 and
SDSSJ1519. Of these four CVs, SDSSJ1232 and SDSSJ1519 display the strong blue continuum 
and weak emission lines consistent with a high accretion rate nova-like system.
The other two objects have strong Balmer lines. SDSSJ1122 is particularly
interesting in that the \ion{He}{2} line is weak but the \ion{He}{1} lines are
unusually strong. Unfortunately, followup observations only could be obtained
for SDSSJ0756+08. 

{\it SDSSJ0756+08}. The SDSS spectrum (Figure 1) resembles that of the IP
SDSSJ0946. Although our coverage is only 1.5 hrs, there is an obvious sinusoidal
modulation apparent indicating a period near 2 hrs. Figure 6 shows the radial 
velocity curves and Table 4 presents the solution with our limited data. If this short period and possible IP nature are confirmed, this object would fall into the small class of IPs with periods under the gap\footnote{see 
http://asd.gsfc.nasa.gov/Koji.Mukai/iphome/iphome.html for a listing of all known IPs}.


\subsection{Systems Showing the Underlying Stars}

As reported in papers I-VII, the systems with lowest accretion rates
allow the underlying stars to be visible, as evidenced by broad Balmer
absorption lines from the white dwarf surrounding the disk emission lines
and/or the presence of TiO bands from a late type secondary. In Figure 1,
SDSSJ0912, SDSSJ1317, SDSSJ1625 and SDSSJ2309 show evidence for an M star. SDSSJ1219 shows the clear signature of a white
dwarf in a low accretion system, and SDSSJ1610 may be in this category. Since
several of the white dwarfs with similar spectra have
been shown to undergo non-radial pulsations, SDSSJ1219 should have
high priority for time-resolved photometry.
SDSSJ0354, SDSSJ1224, SDSSJ1317 and SDSSJ1515 have absorption with very narrow 
emission lines. This group could be pre-CVs with the emission caused by
irradiation from the secondary by a hot white dwarf.

\subsection{Other Disk Systems}

SDSSJ0932 shows a blue continuum with only a broad double-peaked emission
line of H$\alpha$ (Figure 1). Since the photometric magnitudes are similar
in brightness to the spectroscopic fluxes, this is not a dwarf nova in
outburst but likely a high accretion rate disk system (a nova-like variable).

\subsection{ROSAT Correlations}

Six of the 33 systems were detected in X-ray with the 
ROSAT All Sky Survey (RASS; Voges et al. 1999, 2000). 
 Table 5 gives the detected count rates and
exposure times for those CVs. Of the three candidate polars, only
SDSSJ1005 (the possible IP) is detected. It is not too surprising that
all three candidates are not detected as the others could have been in low
accretion states (common for polars) at the time of the ROSAT observations. 

\section{Summary}

The addition of these 33 systems brings the total spectroscopically 
identified CVs in
SDSS to 285. 
 The objects deserving of particular attention are the polar
candidates SDSSJ1207 and SDSSJ1344, the possible IPs SDSSJ0756 and SDSSJ1005,
the object with unusual \ion{He}{1} abundance (SDSSJ1122) and the candidate
for a system containing a pulsating white dwarf (SDSSJ1219).

A summary table of the 285 objects that have SDSS spectra from papers
I-VIII is presented here (Table 6). This table has 22 columns that provide the
object name, RA and Dec in degrees, $u,g,r,i,z$ psf magnitudes and errors,
MJD, plate and fiber of the spectroscopic observation, MJD of the photometric
observation, the spectroscopic target selection flags (detailed in
Stoughton et al. 2002), the period if known, the discovery reference, the 
type of
system and any comments. 
In addition, an active web list of these objects, with links to the spectra and 
information
on types and period is being kept and updated (as new information becomes
available)\footnote{http://www.astro.washington.edu/users/szkody/cvs/index.html}. 
If an outburst is observed, a classification as dwarf nova can
be made, and as spectroscopic/photometric studies appear in the literature, further
orbital periods will become known. At the present time, there are 151 periods
known for these CVs, with 98 (65\%) under the gap ($<$ 2hrs). The list of
objects includes 30 polars (with seven being very low accretion rate polars
or LARPs), six IPs and nine that contain pulsating white dwarfs. While the
SDSS sample is not 100\% complete due to the selection effects discussed
in G\"ansicke et al. (2009), this database will provide a
wealth of information for statistical studies of spectra and characteristics
of CVs for years to come. In addition, future observations of the peculiar
systems which do not seem to fit the prescribed mold of currently known
CVs can provide some useful information on the evolution that occurs in
close binaries.

\acknowledgments

    Funding for the SDSS and SDSS-II has been provided by the Alfred P. Sloan 
Foundation, the Participating Institutions, the National Science Foundation, the
 U.S. Department of Energy, the National Aeronautics and Space Administration, 
the Japanese Monbukagakusho, the Max Planck Society, and the Higher Education 
Funding Council for England. The SDSS Web Site is http://www.sdss.org/.

    The SDSS is managed by the Astrophysical Research Consortium for the 
Participating Institutions. The Participating Institutions are the American 
Museum of Natural History, Astrophysical Institute Potsdam, University of Basel,
 University of Cambridge, Case Western Reserve University, University of 
Chicago, Drexel University, Fermilab, the Institute for Advanced Study, the 
Japan Participation Group, Johns Hopkins University, the Joint Institute for 
Nuclear Astrophysics, the Kavli Institute for Particle Astrophysics and 
Cosmology, the Korean Scientist Group, the Chinese Academy of Sciences (LAMOST),
 Los Alamos National Laboratory, the Max-Planck-Institute for Astronomy (MPIA),
 the Max-Planck-Institute for Astrophysics (MPA), New Mexico State University, 
Ohio State University, University of Pittsburgh, University of Portsmouth, 
Princeton University, the United States Naval Observatory, and the University 
of Washington.

P.S. acknowledges support from NSF grants AST 0607840 and AST 1008734.

\clearpage

\begin{deluxetable}{lccrrrrll}
\tabletypesize{\scriptsize}
\tablewidth{0pt}
\tablecaption{CVs in SDSS}
\tablehead{
\colhead{SDSS J} &  \colhead{MJD-P-F\tablenotemark{a}} & 
\colhead{$g$} & \colhead{$u-g$} & \colhead{$g-r$} & 
\colhead{$r-i$} &
\colhead{$i-z$} & \colhead{P(hr)} & \colhead{Comments\tablenotemark{b}} }
\startdata
034420.16+093006.8 & 54368-2679-299 & 18.19 & -0.28 & 0.00 & 0.03 & 0.00 & ... & \\
035409.32+100924.4 & 54389-2697-028 & 20.89 & 0.38 & -0.09 & -0.03 & -0.16 & ... & \\
073208.11+413008.7 & 54154-2702-411 & 20.77 & -0.31 & 0.32 & 0.20 & 0.39 & ... & DN, W \\
075648.04+305805.0 & 54537-2959-046 & 21.07 & -0.10 & 0.10 & -0.08 & -0.10 & ... & CSS \\
075653.11+085831.8 & 54505-2945-144 & 16.27 & 0.42 & 0.00 & -0.07 & -0.04 & 2 & \\
091242.18+620940.1 & 54450-1786-573 & 18.81 & -0.06 & 0.07 & 0.29 & 0.57 & 2 & DN, W \\
092809.83+071130.5 & 54169-2382-502 & 15.32 & 0.25 & -0.17 & -0.16 & -0.20 & ... & DN, W \\
092918.90+622346.2 & 54465-1787-052 & 18.72 & 0.53 & 0.29 & 0.04 & 0.13 & ... & \\
093220.92+133122.2 & 54093-2578-620 & 18.02 & 0.24 & -0.08 & -0.17 & -0.07 & ...& \\
094634.46+135057.8 & 54139-2582-185 & 17.11 & -0.23 & 0.10 & 0.00 & 0.00 & 4.2 & IP HS 0943+1404 \\
095151.79+471008.7 & 54525-2956-373 & 20.32 & 0.04 & 0.00 & -0.29 & 0.19 & ... & \\
104051.23+151133.6 & 54117-2594-026 & 17.96 & 0.28 & 0.42 & 0.36 & 0.16 & 5.63 & DO Leo \\
100516.61+694136.5 & 54478-1879-608 & 19.41 & 0.32 & 0.48 & 0.31 & 0.34 & ... & IP?, W \\
105550.08+09560.4 & 54498-2886-504 & 19.20 & -0.14 & 0.60 & 0.60 & 0.50 & ... & DN, CSS \\
111703.53+182558.1 & 54453-2857-085 & 14.64 & -0.40 & -0.25 & -0.01 & 0.04 & 42.2 & HK Leo \\
112003.40+663632.4 & 54498-2858-239 & 15.65 & 0.18 & -0.24 & -0.23 & -0.20 & ... & DN, W \\
112253.31$-$111037.6 & 54561-2874-603 & 20.47 & 0.22 & 0.01 & -0.12 & 0.13 & ... & CSS \\
120724.69+223529.8 & 54210-2644-431 & 19.98 & 0.22 & 0.7 & 0.47 & 0.00 & ... & polar? \\
121913.04+204938.3 & 54477-2611-376 & 19.17 & 0.26 & -0.03 & -0.16 & -0.12 & ... & \\
122405.58+184102.7 & 54477-2611-042 & 16.01 & 0.08 & -0.03 & -0.18 & -0.13 & ... & \\
123255.11+222209.4 & 54495-2647-200 & 17.70 & -0.10 & -0.15 & -0.09 & -0.11 & ... & PG1230+226 \\
123932.00+210806.2 & 54481-2613-523 & 18.33 & -0.09 & -0.08 & 0.37 & 0.78 & 2.09 & IR Com \\  
131709.07+245644.2 & 54234-2663-583 & 19.09 & 0.42 & 0.69 & 0.63 & 0.29 & ... & \\
132856.71+310846.0 & 53467-2110-480 & 17.61 & -0.48 & -0.16 & -0.34 & -0.29 & ... & \\
134441.83+204408.3 & 54231-2654-023 & 17.17 & -0.12 & -0.72 & -0.38 & -0.07 & 2 & polar? \\
151500.56+191619.6 & 54271-2793-450 & 17.83 & -0.36 & -0.43 & -0.27 & -0.07 & ... & \\
151915.86+064529.1 & 54562-1834-417 & 16.58 & -0.05 & -0.16 & -0.14 & -0.05 & ... & \\
154539.07+142231.6 & 54567-2517-581 & 16.33 & 0.04 & -0.11 & -0.14 & -0.07 & 4.68 & CT Ser \\
161027.61+090738.4 & 54582-2526-098 & 20.13 & 0.07 & 0.04 & -0.18 & 0.28 & ... & DN, W \\
162207.15+192236.6 & 54589-2970-466 & 18.63 & -0.20 &  0.48 & -0.05 & 0.23 & ... & \\
162520.29+120308.7 & 54572-2531-446 & 18.51 & -0.06 & 0.09 & 0.16 & 0.56 & 2.19 & DN, W \\
163545.72+112458.0 & 54585-2533-156 & 15.21 & 0.32 & -0.13 & -0.11 & -0.10 & 3.4 & V849 Her \\
230949.12+213516.7 & 54328-2623-193 & 16.10 & -0.32 & 0.74 & 0.70 & 0.56 & 4.26 & V405 Peg \\
\enddata
\tablenotetext{a}{MJD-Plate-Fiber for spectra; MJD = JD - 2,400,000.5}
\tablenotetext{b}{DN is a dwarf nova, W is Wils et al. (2010), CSS is Drake
et al. (2009)}
\end{deluxetable}

\clearpage
\begin{deluxetable}{lcccc}
\tablewidth{0pt}
\tablecaption{Follow-up APO Spectra}
\tablehead{
\colhead{SDSSJ} & \colhead{UT Date} &  
\colhead{Time (UT)} & \colhead{Exp (s)} & \colhead{Number of Spectra} }
\startdata
0756+08 & 2010 May 5  & 02:47-04:15 & 600 & 8 \\
0912 & 2009 Mar 29 & 02:20-04:48 & 600 & 12 \\
0929 & 2009 Apr 18 & 03:43-04:38 & 600 & 5 \\
0929 & 2010 May 5  & 04:56-06:39 & 600 & 9 \\
0938\tablenotemark{a} & 2009 Feb 27 & 02:43-07:08 & 600 & 22 \\
1055 & 2010 May 5  & 04:31 & 600 & 1 \\
1344 & 2009 Apr 18 & 05:10-06:56 & 600 & 9 \\
\enddata
\tablenotetext{a}{object discovered in Paper VII}
\end{deluxetable}

\clearpage
\begin{deluxetable}{lrrrrrrrr}
\tabletypesize{\scriptsize}
\tablewidth{0pt}
\tablecaption{SDSS Emission Line Fluxes and Equivalent Widths\tablenotemark{a}}
\tablehead{
\colhead{SDSSJ} &  
\multicolumn{2}{c}{H$\beta$} &
\multicolumn{2}{c}{H$\alpha$} &
\multicolumn{2}{c}{He4471} & \multicolumn{2}{c}{HeII4686}\\
\colhead{} &  \colhead{F} &
\colhead{EW} & \colhead{F} & \colhead{EW} & \colhead{F} & \colhead{EW} &
\colhead{F} & \colhead{EW} } 
\startdata
0344 & 2.7 & 10 & 2.9 & 16 & 0.7 & 2 & ... & ... \\
0354 & 0.1 & 9.5 & 0.4 & 53 & ... & ... & ... & ... \\
0732 & 0.6 & 23 & 0.8 & 29 & ... & ... & ... & ... \\
0756+30 & 0.9 & 69 & 1.1 & 136 & 0.2 & 17 & ... & ... \\
0756+08 & 11.7 & 23 & 10.6 & 35 & 4.1 & 7 & 9.9 & 18.5 \\
0912 & 2.6 & 44 & 4.9 & 124 & ... & ... & ... & ... \\
0928 & ... & ... & ... & ... & ... & ... & ... & ... \\ 
0929 & 1.4 & 9 & 1.6 & 14 & 0.5 & 3 & 0.6 & 3.6 \\
0932 & ... & ... & 1.1 & 8.5 & ... & ... & ... & ... \\
0946 & 5.7 & 34 & 14.2 & 48 & 4.4 & 8 & 12.7 & 24 \\
0951 & 1.7 & 76 & 2.5 & 243 & ... & ... & ... & ... \\
1005 & 7.7 & 43 & 6.2 & 46 & 2.1 & 11 & 5.1 & 29 \\
1040 & 13.5 & 4.3 & 23.3 & 15 & ... & ... & 21.0 & 6 \\
1055 & 14.9 & 3 & 22.8 & 9 & ... & ... & ... & ... \\
1117 & 23.3 & 4 & 24.5 & 9 & 5.3 & 0.6 & 11.8 & 2 \\
1120 & 0.8 & 0.9 & 10 & ... & ... & ... & ... \\
1122 & 0.5 & 24 & 0.9 & 87 & 0.3 & 10 & 0.1 & 4 \\
1207 & 1.2 & 35 & 1.2 & 45 & 0.5 & 16 & 0.7 & 20 \\
1219 & 1.6 & 22 & 3.7 & 87 & ... & ... & ... & ... \\
1224 & 5.6 & 4 & 6.4 & 9 & ... & ... & ... & ... \\
1232 & 1.7 & 5 & 2.3 & 13 & ... & ... & 1.2 & 3 \\
1239 & 41.6 & 106 & 59.1 & 222 & ... & ... & ... & ... \\
1317 & 1.2 & 12 & 1.8 & 13 & 0.2 & 2 & ... & ... \\
1328 & 15.3 & 33 & 49.7 & 142 & 1.2 & 2 & ... & ... \\
1344 & 1.1 & 7 & 1.4 & 19 & ... & ... & ... & ... \\
1515 & 1.2 & 3 & 1.0 & 6 & ... & ... & ... & ... \\
1519 & 4.3 & 4 & 7.3 & 16 & ... & ... & 2.0 & 2 \\
1545 & 5.7 & 5 & 6.2 & 10 & 1.1 & 0.7 & 4.1 & 3 \\
1610 & 0.6 & 19 & 1.3 & 76 & ... & ... & ... & ... \\
1622 & 14.1 & 100 & 12.9 & 111 & 3.1 & 20 & 0.8 & 6 \\
1625 & 1.6 & 33 & 3.5 & 111 & ... & ... & ... & ... \\
1635 & ...& ... & 3.5 & 2 & ... & ... & ... & \\
2309 & 85.4 & 82 & 122.6 & 71 & 15.4 & 16 & ... & ... \\
\enddata
\tablenotetext{a}{Fluxes are in units of 10$^{-15}$ ergs cm$^{-2}$ s$^{-1}$,
equivalent widths are in units of \AA}
\end{deluxetable}

\clearpage
\normalsize
\begin{deluxetable}{lcccccc}
\tablewidth{0pt}
\tablecaption{Radial Velocity Solutions}
\tablehead{
\colhead{SDSSJ} & \colhead{Line} & \colhead{P (min)\tablenotemark{a}} & 
\colhead{$\gamma$ (km s$^{-1}$)} & \colhead{K (km s$^{-1}$)} &
\colhead{T$_{0}$ (UT)} & \colhead{$\sigma$ (km s$^{-1}$)} }
\startdata
0756+08 & H$\alpha$ & 117 & 241$\pm$5 & 166$\pm$16 & 4:15 & 24 \\
0756+08 & H$\beta$ & 110 & 192$\pm$7 & 129$\pm$28 & 4:11 & 44 \\
0912 & H$\alpha$ & 113\tablenotemark{b} & 1.1$\pm$2.5 & 40$\pm$8 & 4:04 & 17 \\
0912 & H$\beta$ & 113 & 22$\pm$3 & 60$\pm$9 & 4:21 & 18 \\
1344 & H$\alpha$ & 122 & 34$\pm$1 & 374$\pm$15 & 5:32 & 27 \\
1344 & H$\beta$ & 110 & 47$\pm$2 & 412$\pm$23 & 5:33 & 42 \\
\enddata
\tablenotetext{a}{Periods are generally uncertain by 
10\%, as evidenced by the dispersion between values obtained from the two 
lines.} 
\tablenotetext{b}{Due to low K and large sigma, best fit provided by fixing
the period to that determined from H$\beta$.}
\end{deluxetable}

\clearpage
\begin{deluxetable}{lccll}
\tablewidth{0pt}
\tablecaption{ROSAT Detections}
\tablehead{
\colhead{SDSSJ} & \colhead{ROSAT (c s$^{-1}$)\tablenotemark{a}} & \colhead{Exp (s)}
& \colhead{RXS} & \colhead{Type} }
\startdata
0946 & 0.027$\pm$0.011 & 345 & RXS J094635.9+135126 & IP \\
0951 & 0.0076$\pm$0.0020 & 2712 & RXP J095152.1+471004 & ... \\
1005 & 0.048$\pm$0.018 & 243 & RXS J100518.3+694101 & IP? \\
1117 & 0.0023$\pm$0.0006 & 23865 & WGA J1117.0+1826=HK Leo & ... \\
1239 & 0.061$\pm$0.013 & 431 & RXS J123930.6+210815=IR Com & DN \\
2309 & 0.212$\pm$0.022 & 466 & RXS J230949.6+213523=V405 Peg & NL \\
\enddata
\tablenotetext{a}{For a 2 keV bremsstrahlung spectrum, 1 c s$^{-1}$ corresponds to a 
0.1-2.4 keV flux of about 7$\times10^{-12}$ ergs cm$^{-2}$ s$^{-1}$.}
\end{deluxetable}

\clearpage
\begin{deluxetable}{lll}
\tablewidth{0pt}
\tablecaption{SDSS I/II CV Catalog Format\tablenotemark{a}}
\tablehead{
\colhead{Column} & \colhead{Format} & \colhead{Description} }
\startdata
1 & A18 & SDSS Designation hhmmss.ss+ddmmss.s (J2000) \\
2 & F11.6 & R. A. (J2000) in degrees \\
3 & F11.6 & Dec. (J2000) in degrees \\
4 & F7.3 & u PSF magnitude \\
5 & F6.3 & error in u PSF magnitude \\
6 & F7.3 & g PSF magnitude \\
7 & F6.3 & error in g PSF magnitude \\
8 & F7.3 & r PSF magnitude \\
9 & F6.3 & error in r PSF magnitude \\
10 & F7.3 & i PSF magnitude \\
11 & F6.3 & error in i PSF magnitude \\
12 & F7.3 & z PSF magnitude \\
13 & F6.3 & error in z PSF magnitude \\
14 & I6 & MJD of spectroscopic observation \\
15 & I5 & Plate of spectroscopic observation \\
16 & I4 & Fiber of spectroscopic observation \\
17 & I6 & MJD of photometric observation \\
18 & I12 & Spectrscopic target selection flag \\
19 & F6.2 & Period (hours); 0.00 is unknown period \\
20 & I3 & Reference \\
21 & A6 & Type of CV System \\
22 & A40 & Comments \\
\enddata
\tablenotetext{a}{Table 6 is published in its entirety in the electronic
edition of AJ}
\end{deluxetable}

\clearpage
\begin{figure} [p]
\figurenum {1a}
\plotone{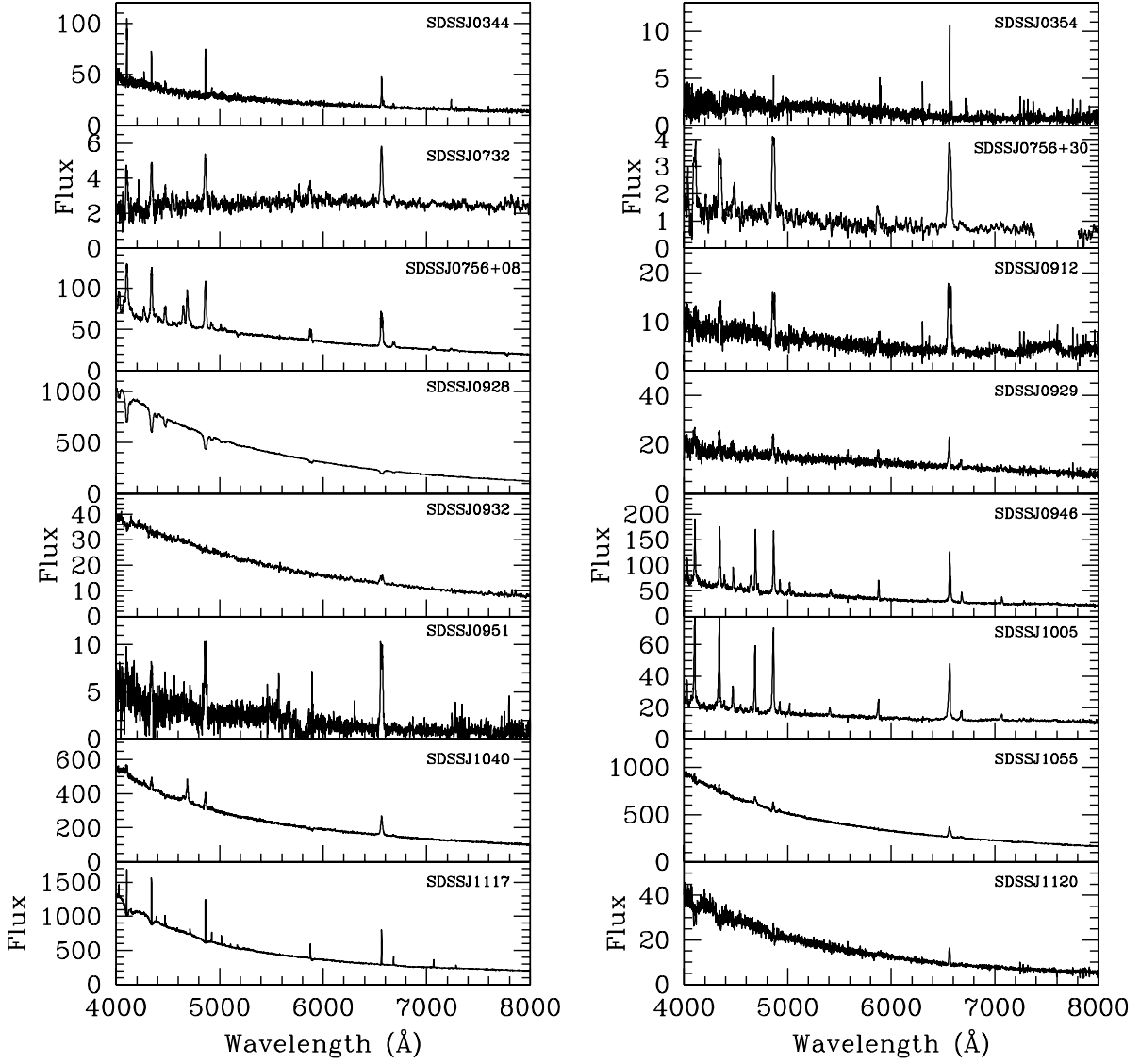}
\caption{SDSS spectra of the 33 CVs.
 Vertical axis is 
units of
flux density F$_{\lambda}\times$10$^{-17}$ ergs cm$^{-2}$ s$^{-1}$ \AA$^{-1}$. The 
spectral
resolution is about 3\AA. The gaps in SDSSJ0756+30 (between 7400-7800\AA) and 
SDSSJ1239 (between 4000-4800\AA) are due to unusable sections of 
the data.}
\end{figure}

\clearpage

\begin{figure} [p]
\figurenum {1b}
\plotone{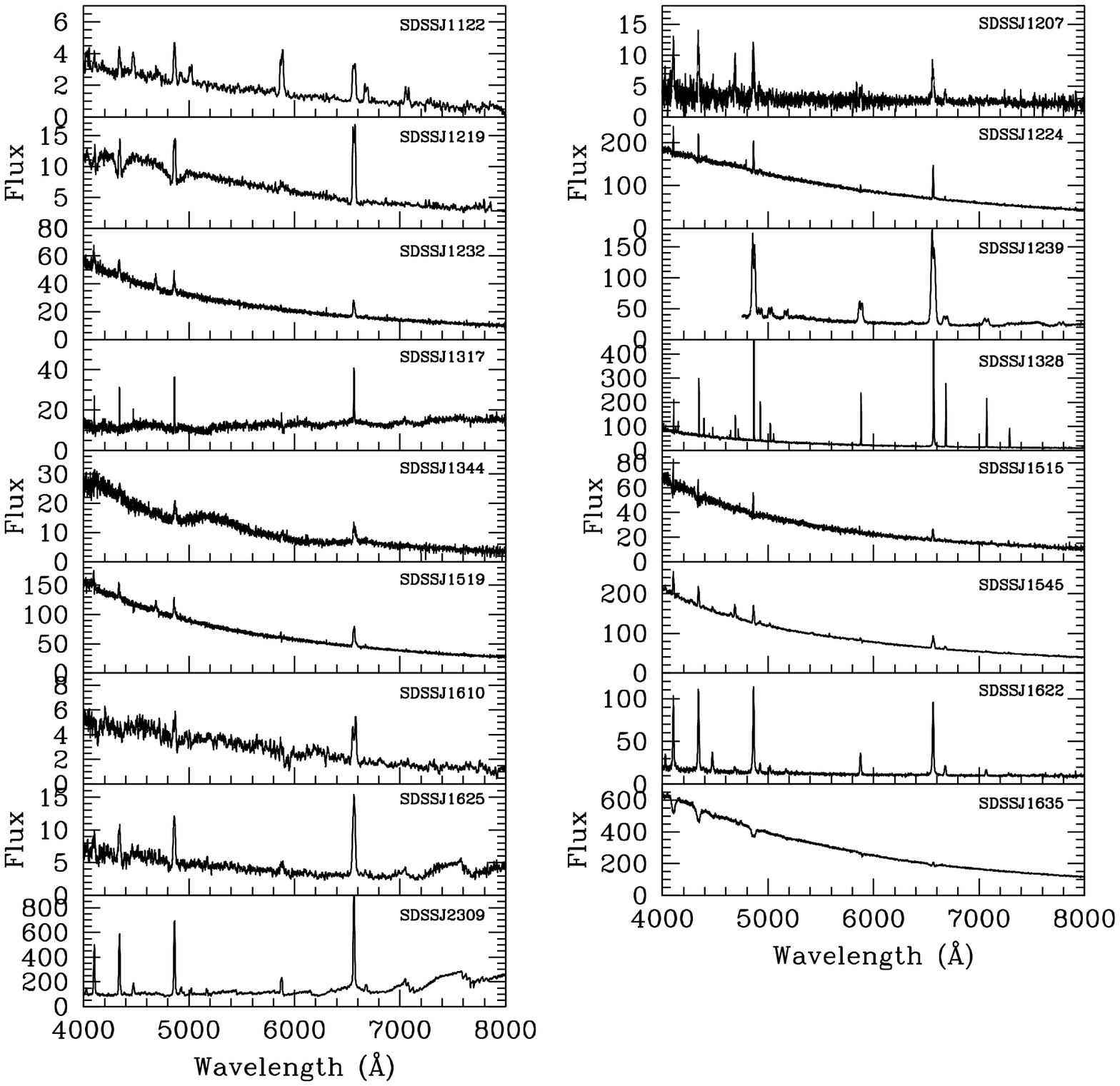}
\caption{Figure 1 continued.}
\end{figure}

\begin{figure} [p]
\figurenum {2}
\plotone{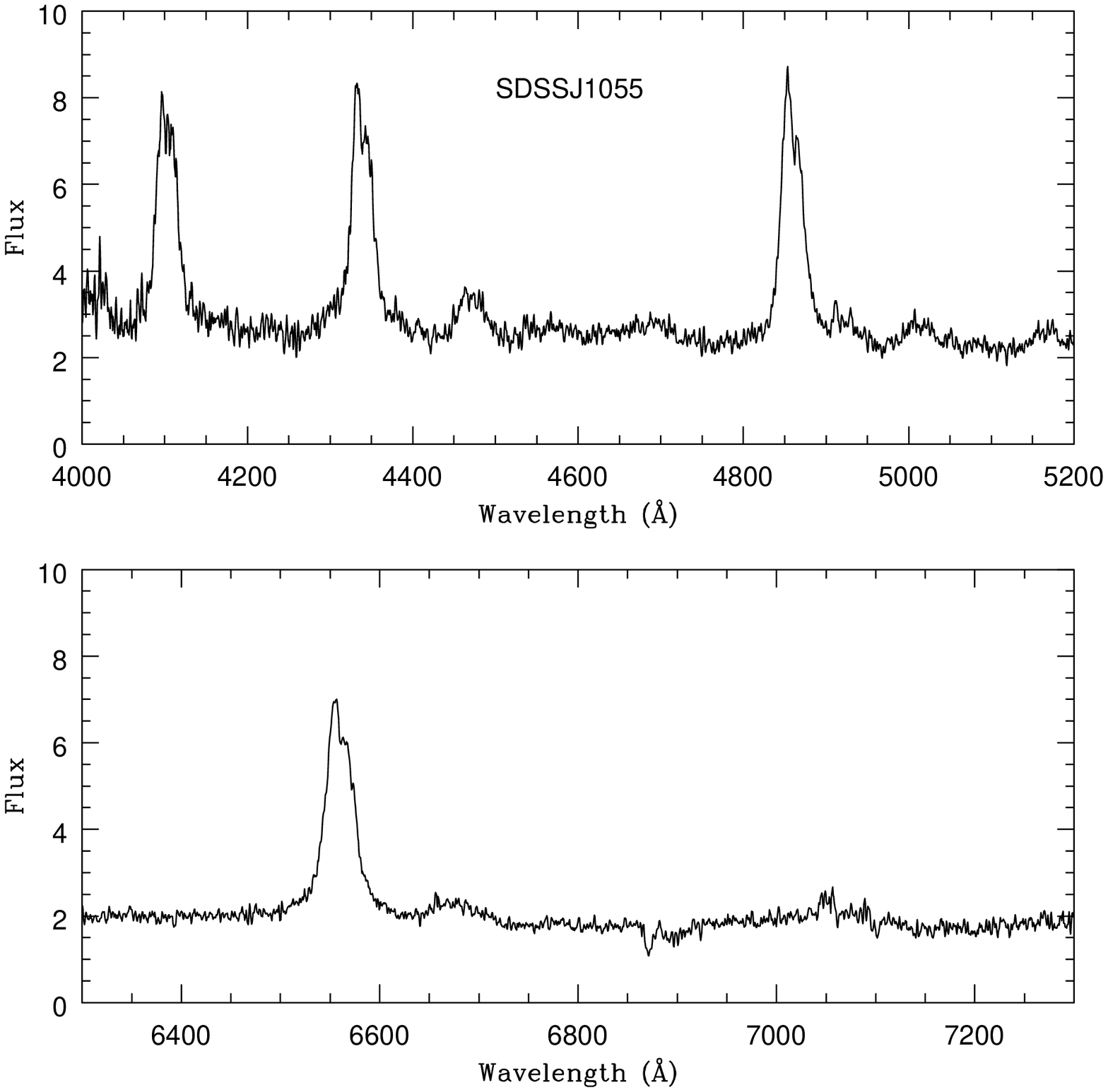}
\caption{APO blue and red spectra of SDSSJ1055 at quiescence. Vertical
axis is flux density F$_{\lambda}\times$10$^{-16}$ ergs cm$^{-2}$ s$^{-1}$ \AA$^{-1}$}
\end{figure}

\begin{figure} [p]
\figurenum {3}
\plotone{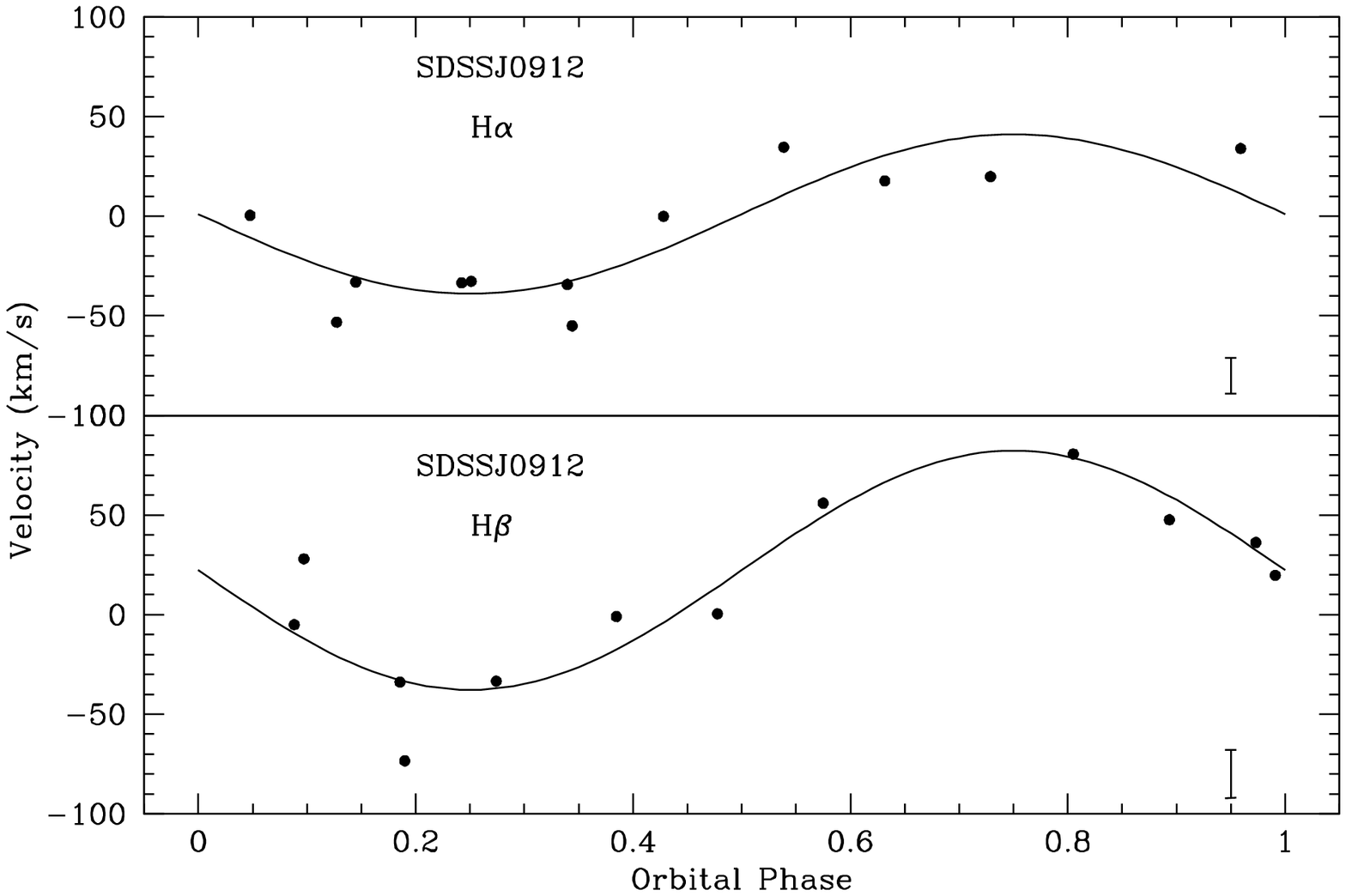}
\caption{H$\alpha$ and H$\beta$ velocity curves of SDSSJ0912 from the APO
data. The smooth curves are the 
best fit sinusoids (Table 4). Sigmas of the sine fits are listed in Table 4.
Estimates of the observed error on the measured velocities is shown by the
error bars near phase 0.9.}
\end{figure}

\begin{figure} [p]
\figurenum {4}
\epsscale{0.8}
\plotone{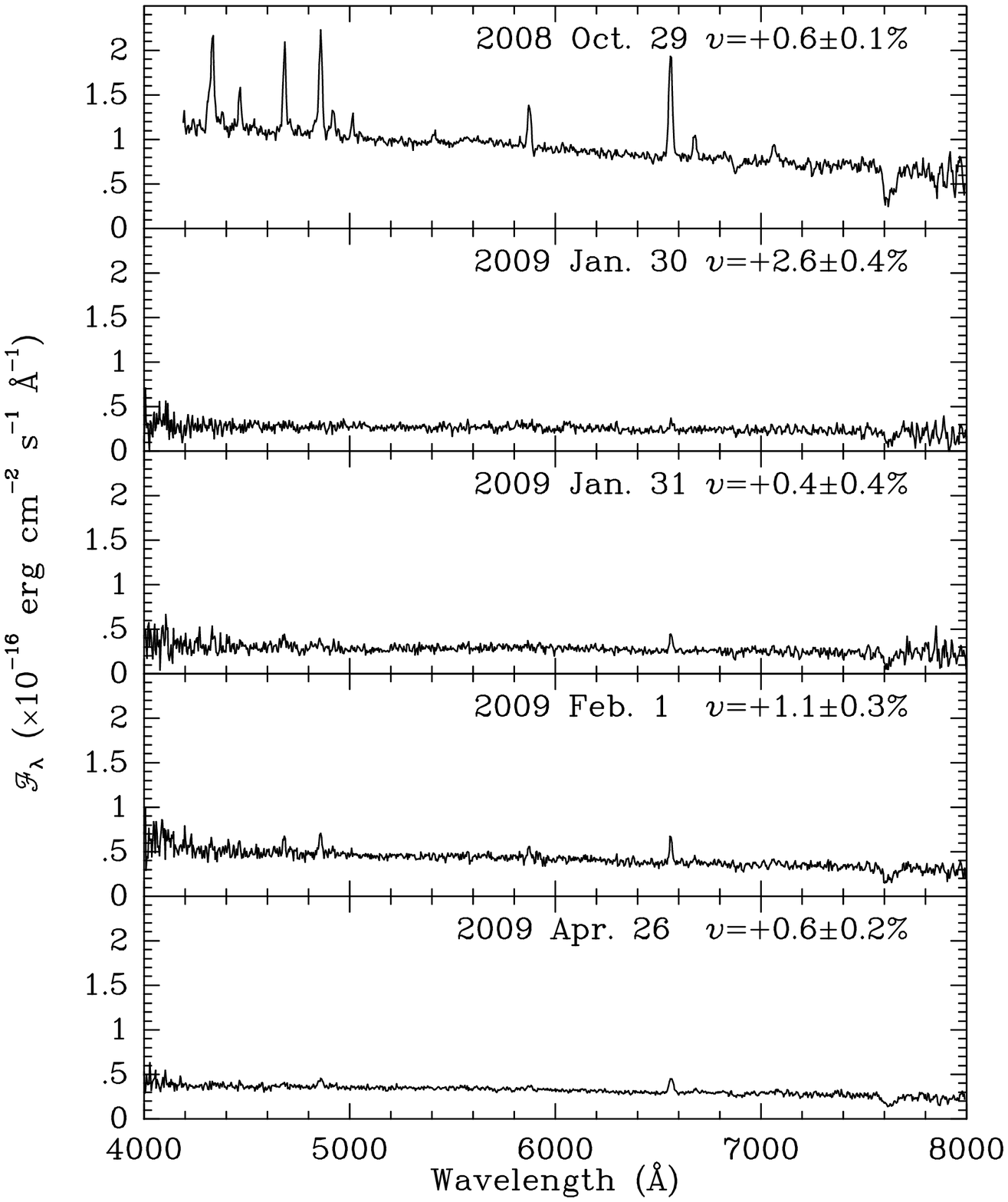}
\vspace{0.5in}
\caption{Spectra and polarization values of SDSSJ0938 for 5 nights in
2008-2009 during high and low states.}
\end{figure}

\begin{figure} [p]
\figurenum {5}
\plotone{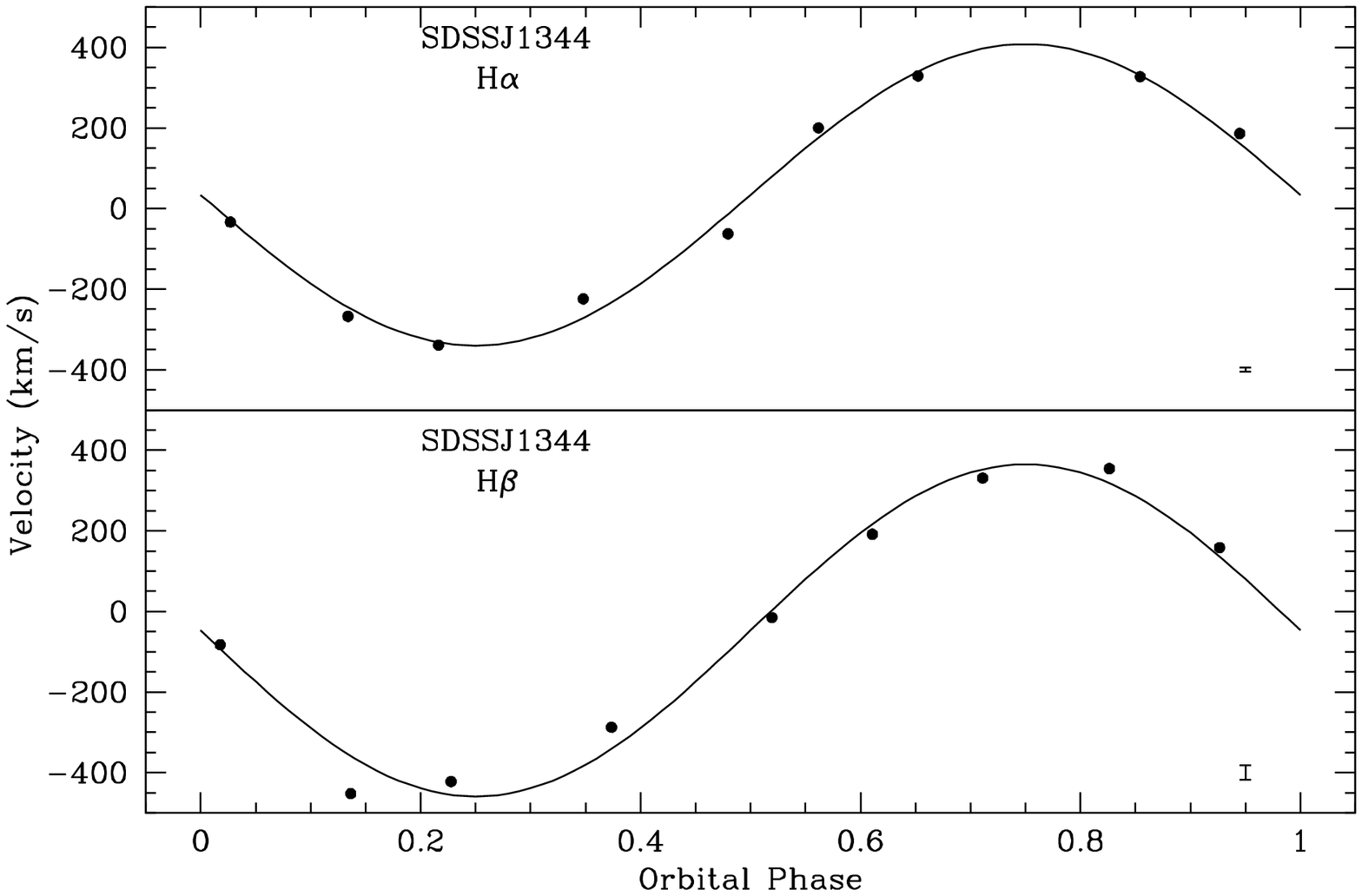}
\caption{H$\alpha$ and H$\beta$ velocity curves of SDSSJ1344 from APO data with
the best fit sinusoids from Table 4 superposed. Estimates of the measured
velocities are shown by the error bars near phase 0.9.}
\end{figure}

\begin{figure} [p]
\figurenum {6}
\plotone{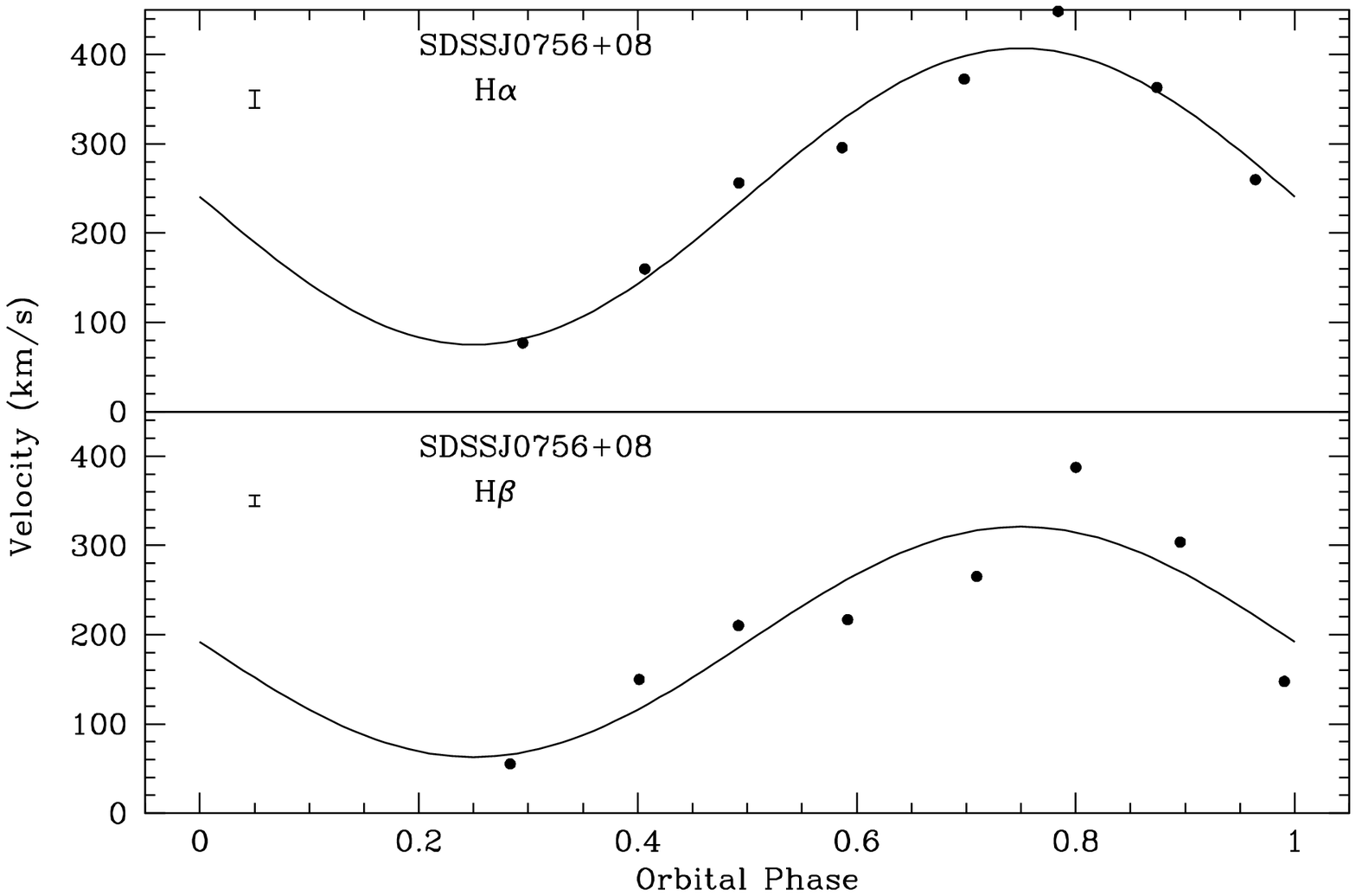}
\caption{H$\alpha$ and H$\beta$ velocity curves of SDSSJ0756+08 from
APO data with
the best fit sinusoids from Table 4 superposed. Estimates of the error on
the measured velocities are shown near phase 0.05.}
\end{figure}

\end{document}